\documentclass[fleqn,usenatbib]{mnras}

\usepackage{newtxtext,newtxmath}

\usepackage[T1]{fontenc}
\usepackage{ae,aecompl}
\usepackage{ulem} 

\usepackage{color}

\usepackage{graphicx}	
\usepackage{amsmath}	
\usepackage{amssymb}	

\title[Maximum stellar surface density of galaxies]{What determines the maximum stellar surface density of galaxies?}

\author[Ling et al.]{Chih-Teng Ling$^{1}$\thanks{E-mail: s106022129@m106.nthu.edu.tw}, Tetsuya Hashimoto$^{2,3}$, Tomotsugu Goto$^{2}$, Ting-Yi Lu$^{2}$, 
\newauthor
Alvina Y. L. On$^{2,3,4}$, Daryl Joe D. Santos$^{2}$,  Tiger Y.-Y. Hsiao$^{5,6}$ and Simon C. -C. Ho$^{2}$
\\
\\
$^{1}$Department of Physics, National Tsing Hua University, 101, Section 2. Kuang-Fu Road, Hsinchu, 30013, Taiwan (R.O.C.)\\
$^{2}$Institute of Astronomy, National Tsing Hua University, 101, Section 2. Kuang-Fu Road, Hsinchu, 30013, Taiwan (R.O.C.)\\
$^{3}$Centre for Informatics and Computation in Astronomy (CICA), National Tsing Hua University, 101, Section 2. Kuang-Fu Road, Hsinchu, 30013, Taiwan (R.O.C.)\\
$^{4}$Mullard Space Science Laboratory, University College London, Holmbury St Mary, Surrey RH5 6NT, UK\\
$^{5}$Department of Atmospheric Science, National Central University, No.300, Zhongda Rd., Zhongli Dist., Taoyuan City 32001, Taiwan (R.O.C.)\\
$^{6}$Institute of Astronomy and Astrophysics, Academia Sinica, Taipei 10617, Taiwan (R.O.C.)
}

\date{Accepted 2020 May 26. Received 2020 April 9; in original form 2019 August 8}

\pubyear{2020}

\begin{document}
\label{firstpage}
\pagerange{\pageref{firstpage}--\pageref{lastpage}}

\maketitle

\begin{abstract}
Observationally, it has been reported that the densest stellar system in the Universe does not exceed a maximum stellar surface density, $\Sigma^{\max}_{*}$ = $3\times10^5$M\textsubscript{\(\odot\)}pc$^{-2}$, throughout a wide physical scale ranging from star cluster to galaxy.
This suggests there exists a fundamental physics which regulates the star formation and stellar density.
However, factors that determine this maximum limit are not clear.
In this study, we show that $\Sigma^{\max}_{*}$ of galaxies is not a constant as previous work reported, but actually depends on the stellar mass.
We select galaxy sample from the Sloan Digital Sky Survey Data Release 12 at $z=0.01-0.5$.
In contrast to a constant maximum predicted by theoretical models, $\Sigma^{\max}_{*}$ strongly depends on stellar mass especially for less massive galaxies with $\sim10^{10}$M\textsubscript{\(\odot\)}.
We also found that a majority of high-$\Sigma_{*}$ galaxies show red colours and low star-formation rates. These galaxies probably reach the $\Sigma^{\max}_{*}$ as a consequence of the galaxy evolution from blue star forming to red quiescent by quenching star formation.
One possible explanation of the stellar-mass dependency of $\Sigma^{\max}_{*}$ is a mass dependent efficiency of stellar feedback.
The stellar feedback could be relatively more efficient in a shallower gravitational potential, which terminates star formation quickly before the stellar system reaches a high stellar density.

\end{abstract}

\begin{keywords}
galaxies: evolution $-$ galaxies: fundamental parameters
\end{keywords}



\section{Introduction}
An existence of a maximum value of stellar surface density, $\Sigma_{*}^{\rm max}$, was proposed by \cite{Hopkins2010}.
Observationally, in any stellar system in the Universe including star clusters and galaxies, the stellar surface density does not exceed $\Sigma_{*}^{\rm max}$ $\sim 3\times 10^{5}M_{\odot}$ pc$^{-2}$\citep{Hopkins2010,Grudic2019}.
Extremely compact massive early-type galaxies has been discovered at redshift $\sim 2$ or even higher \citep[e.g.,][]{vanDokkum2008,2017Nature,QGs2018,Kubo2018}.
Such extreme galaxies are even less dense than $\Sigma_{*}^{\rm max}$ \citep{Kubo2018}, even though galaxy size evolution models suggest the denser galaxies at the higher redshift \citep[e.g.,][]{Bezanson2009,Naab2009}.

These observational evidences suggest a universal physics that controls star formation and its surface density, spanning up to $\sim$ 8 order of magnitude in stellar mass as well as a wide redshift range.
\cite{Hopkins2010} argued that a key physics to determine $\Sigma_{*}^{\rm max}$ is feedback from massive stars.
As in the case of black hole accretion, a strong feedback from intense star formation could suppress further gas accretion and then limits the final stellar density.
However, the strength of the self-gravity actually overcomes feedback when the gas cloud collapsed as expected from simulations and a simple analytic model \citep{Grudic2018}. 
This is because the strength of the self-gravity is proportional to ($M$/$R$)$^{2}$ while feedback is proportional to $M$ \citep{Grudic2018}, where $R$ and $M$ are the size and total mass of the gas cloud, respectively.
Therefore, the relative strength of the self-gravity to feedback is determined by the surface density.
As a consequence, the star-forming efficiency (SFE), that is defined by star-formation rate (SFR) divided by molecular gas mass, increases as the gas cloud contracts \citep{Grudic2018} against the strong feedback scenario of massive stars.

\cite{Grudic2019} proposed a new model to address the maximum limit of surface stellar density based on a quenching of the star formation due to a gas consumption rather than the strong feedback. 
They constructed a model of a gas conversion into stars with the stellar feedback assuming a surface-density dependence of SFE that is confirmed in the simulations \citep{Grudic2018}.
According to the model, SFE dramatically increases when the surface density exceeds a critical surface density, $\Sigma_{\rm crit}$, and converges into a maximum SFE, $\epsilon_{ff}^{\rm max}$, as the gas cloud collapses.
The high SFE rapidly consumes the remaining gas and then the star formation terminates before the system reaches the maximum surface density.
The model successfully explains the existence of the maximum surface density by assuming a reasonable parameter range of $\epsilon_{ff}^{\rm max}$. 

In this paper, we investigate the maximum surface density of galaxies selected from the Sloan Digital Sky Survey (SDSS) Data Release 12 \citep{Alam2015}, especially on how the maximum value depends on a galaxy stellar mass which is one of the most fundamental physical parameters of galaxies.
The structure of the paper is as follows.
In Section \ref{data_selection}, we describe selection criteria of our galaxy sample.
Derivation of physical quantities of the sample is demonstrated in Section \ref{analysis}.
In Section \ref{result}, high surface density galaxies are specified and we demonstrate $\Sigma_{*}^{\rm max}$ is actually not constant but stellar-mass dependent in contrast to the model prediction.
We discuss possible physical qualitative interpretations of the stellar-mass dependency of $\Sigma_{*}^{\rm max}$ in Section \ref{discussion} followed by a conclusion in Section \ref{conclusion}.

Throughout the paper, we assume $\Lambda$CDM cosmology with ($\Omega_{M}$, $\Omega_{\Lambda}$, $\Omega_{b}$, $h_0) = ($0.307, 0.691, 0.048, 67.7) and a flat universe assumption unless otherwise mentioned \citep{Planck2015_2016A&A...594A..13P}.

\section{Data selection}
\label{data_selection}
Galaxies were selected from the Sloan Digital Sky Survey Data Release 12 \citep[SDSS DR12;][]{Alam2015}.
We used a stellar-mass catalogue from the Wisconsin Group.
The stellar mass was calculated by the Principal Component Analysis (PCA) Method \citep{Chen2012} using the \citet{Maraston2011} stellar population synthesis models.
In this work, we used the SDSS SQL table columns of PhotoObjAll, stellarMassPCAWiscM11, and emissionLinesPort for photometries, galaxy stellar mass, and emission-line fluxes, respectively.

To ensure that our sample only contains galaxies with reliable measurement, objects which had non-zero warning value in the table were excluded.
We also applied CLEAN photometric flag to our sample to exclude most of the suspicious and duplicate objects.
Petrosian magnitudes were adopted to calculate galaxy colours for our sample.
Due to the SDSS spectroscopic $r$-band flux limit, we selected only galaxies brighter than 17.77 mag in $r$-band \citep{Strauss2002}.
The redshift range of the sample was limited to $0.01-0.5$, because at $z>$ 0.5, the fraction of normal galaxies significantly decreases in the SDSS, and the quality of the SDSS spectra becomes poorer.
We avoided galaxies at $z<0.01$ because the redshift uncertainty due to the peculiar velocity is large at such low redshift \citep[e.g.,][]{2013A&A...557A..21S}, so are the derived luminosity and stellar mass. In addition,  $z<0.01$ are contaminated by saturated stars disguised as galaxies (extended objects). 

\begin{figure}
	\includegraphics[width=.52\textwidth]{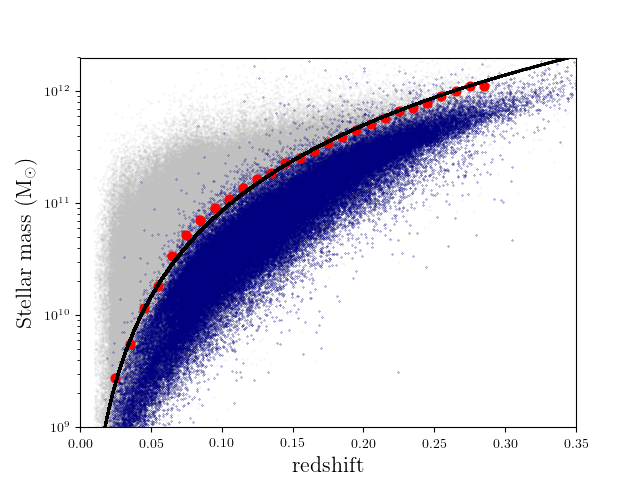}
    \caption{
    The stellar-mass completeness as a function of redshift. 
    Grey dots: all galaxies; blue dots: 20\% faintest galaxies; red dots: 2$\sigma$ above the stellar-mass median of the faintest galaxies subset in each redshift bins; black line: the stellar-mass completeness function as the logarithmic fitting of $2\sigma$ (red dots). Galaxies located below the black line are excluded.
    }
    \label{fig1}
\end{figure}

The completeness of the sample at the low-mass end has to be treated carefully due to the flux limit of SDSS ($r$-band magnitude = 17.77) which causing redder galaxies to be dropped from the sample.
To exclude this bias, we follow \cite{2010A&A...523A..13P} to calculate the stellar-mass completeness function.
We first selected galaxies with r-band Petrosian magnitudes between 17.6 and 17.77 mag in our sample (hereafter 20\% faintest galaxies), and divided them into bins of 0.1 redshift.
We calculated the median and $\sigma$ of the stellar masses in each redshift bin.
The stellar-mass completeness function is then determined by fitting a logarithmic function to the median$+2\sigma$. 
Figure \ref{fig1} illustrates our selection and fitting of stellar-mass completeness function.
Hence we have limited our sample to the stellar-mass completeness line and above. This ensures our sample is at 95\% completeness level, even at the lowest stellar mass limit.

In summary, a total of 185,688 galaxies were selected for our analysis based on the following criteria.
\begin{itemize}
\item 0.01 $<$ redshift $<$ 0.5
\item $r$-band Petrosian magnitude $<17.77$
\item stellar mass $>$ stellar-mass completeness function
\item CLEAN = 1 
\item (calibStatus\_r \& 1) $\neq 0$
\item warning $=0$
\item zWarning $=0$
\end{itemize}

\section{Analysis}
\label{analysis}
The SDSS DR12 sources include two independent galaxy size measurements along with the likelihoods assuming exponential disk or de Vaucouleurs profile.
Since SDSS did not provide a multi-component fitting, we select the single component fitting with higher likelihoods for each galaxy in our sample.
For two-component galaxies, the single component fitting results in biases of $\sim 5-10\%$ in radius \citep{10.1093/mnras/stt822}, which is acceptable for our purpose.

H$\alpha$ luminosity is calculated from the H$\alpha$ flux in the SDSS DR12 catalogue and luminosity distance to the individual sources.
The H$\alpha$ luminosity is converted to star formation rate (SFR) assuming the global Schmidt law by \citet{Kennicutt1998}.
We used $k$-correction \citep{Chilingarian2010,Chilingarian2012} to calculate the absolute Petrosian magnitudes in all bands and  the rest frame $g-r$ colour.

The stellar surface density is defined as
\begin{eqnarray}
\Sigma_{*} = \frac{M_{*}}{\pi R^2},
\end{eqnarray}
where $M_{*}$ is the stellar mass ($M_{\odot}$) and $R$ is the size of galaxy in parsec which we have calculated from apparent radius and redshift.

\begin{figure}
	\includegraphics[width=.52\textwidth]{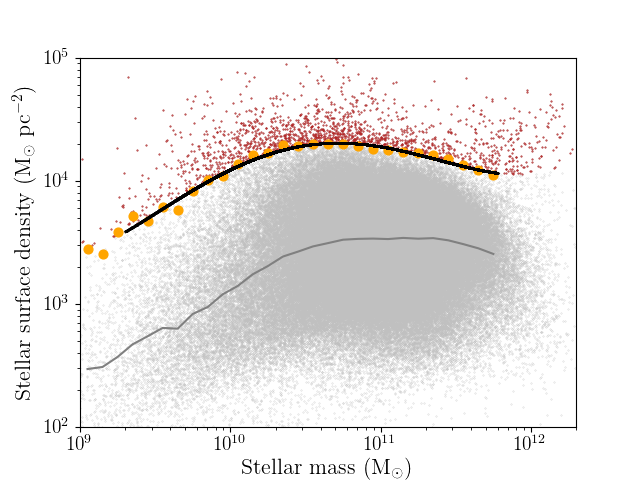}
    \caption{
    Stellar surface density, $\Sigma_{*}$, as a function of stellar mass.
    Orange dots: $2\sigma$ above $\Sigma_{\rm median}$ of subsamples in each stellar mass bin; grey dots: all galaxies in our sample; red dots: high-$\Sigma_{*}$ galaxies; grey line: $\Sigma_{\rm median}$ line;  black line: polynomial fit of $2\sigma$ data (orange dots).
    }
    \label{fig3}
\end{figure}

Here we define high-$\Sigma_{*}$ galaxies as those with stellar surface densities more than $2\sigma$ higher than the median value ($\Sigma_{*}^{\rm median}$).
We also re-define $\Sigma^{\max}_{*}$ as the stellar surface densities of high-$\Sigma_{*}$ galaxies in this work.
$\Sigma^{\max}_{*}$ in our sample was calculated as follows.
First, the sample was divided into subsamples with different stellar mass bins.
The stellar mass bin is 0.1 dex in logM$_{*}$ from 10$^{9}$ to 10$^{12}$ M$_{\odot}$.
For each bin, we derived medians and standard deviations of $\Sigma_{*}$.
The median and 2$\sigma$ value in each bin were fitted with fourth degree polynomial functions.
In Figure \ref{fig3}, the defined high-$\Sigma_{*}$ galaxies are shown by red dots with the background full sample in grey dots.
The best fit lines of $\Sigma^{\max}_{*}$ criterion and median distribution are displayed by black and grey solid lines, respectively.
The high-$\Sigma_{*}$ samples include 1,743 galaxies, which is about $1\%$ of our total galaxies sample.

\section{Result}
\label{result}
In Figure \ref{fig3} we find that the $\Sigma^{\max}_{*}$ increases with stellar mass, especially in low mass range at $10^{9.5}$M$_{\odot}-10^{10.5}$M$_{\odot}$.
This indicates that $\Sigma^{\max}_{*}$ is dependent on stellar mass in contrast to a proposed constant $\Sigma^{\max}_{*}$ by \citet{Grudic2019}.
Besides, Figure \ref{fig3} shows that the $\Sigma^{\max}_{*}$ of galaxy peaks at M$_{*}$ $\sim10^{10.5}$M$_{\odot}$.
We find the $\Sigma^{\max}_{*}$ flattens and even decreases at M$_{*}$ $>10^{10.5}$M$_{\odot}$.

\begin{figure}
	\includegraphics[width=.52\textwidth]{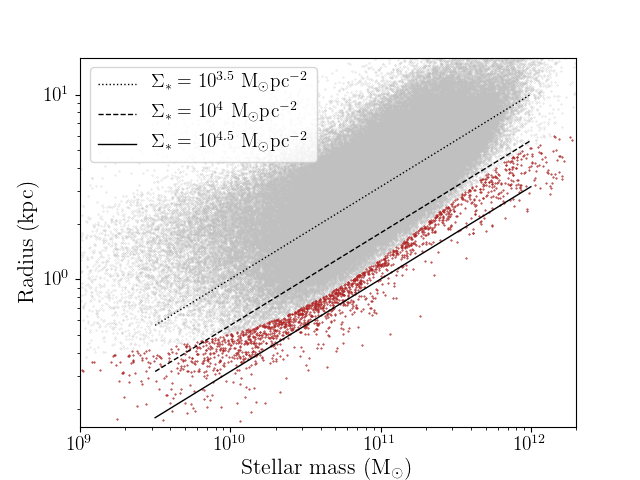}
    \caption{
    Radius of high-$\Sigma_{*}$ galaxies as a function of stellar mass.
    Grey dots: all galaxies; red dots: high-$\Sigma_{*}$ galaxies.
    Black dotted, dashed, and solid lines correspond to $\Sigma_{*}$=10$^{3.5}$, 10$^{4.0}$, and $10^{4.5} M_{\odot}$ pc$^{-2}$, respectively.
    }
    \label{fig4}
\end{figure}

Figure \ref{fig4} shows the radius of the high-$\Sigma_{*}$ galaxies (red dots) as a function of stellar mass.
The radius of high-$\Sigma_{*}$ galaxies keeps in a constant value from $\sim$10$^{9.5}$ M$_{\odot}$ to $\sim10^{10.5}$M$_{\odot}$.
The radius increase with increasing stellar mass from $\sim10^{10.5}$M$_{\odot}$.
The slope of the increasing radius is steeper than that of a constant stellar surface density, which results in the turn over of $\Sigma^{\max}_{*}$ at $>$10$^{10.5}$ M$_{\odot}$ in Figure \ref{fig3}.
The $\Sigma^{\max}_{*}$ of our galaxy sample is $\sim 5\times 10^4 $M$_{\odot}$pc$^{-2}$ at M$_{*}\sim 10^{10.5}$M$_{\odot}$.
While the $\Sigma^{\max}_{*}$ peaks at M$_{*}$ $\sim 10^{10.5}$M$_{\odot}$, there are some more massive galaxies that have such high maximum surface density ($\sim5\times 10^4 $M$_{\odot}$pc$^{-2}$).

\begin{figure}
	\includegraphics[width=.52\textwidth]{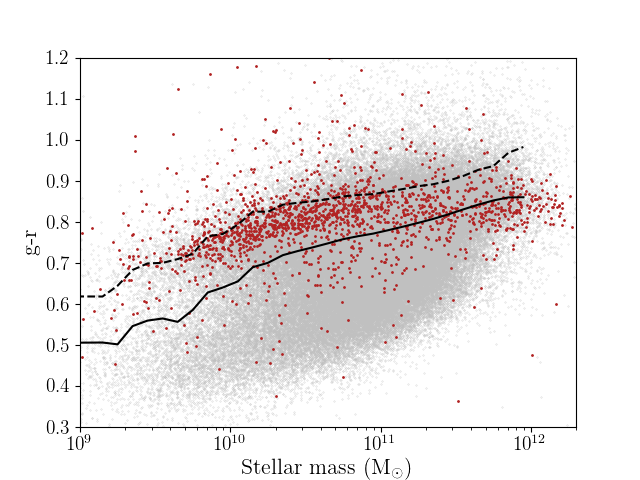}
    \caption{
    $g-r$ colour as a function of stellar mass.
    Grey dots: all galaxies; red dots: high-$\Sigma_{*}$ galaxies; solid line: median line; dashed line: median$+\sigma$ line.
    }
    \label{fig5}
\end{figure}

\begin{figure}
	\includegraphics[width=.52\textwidth]{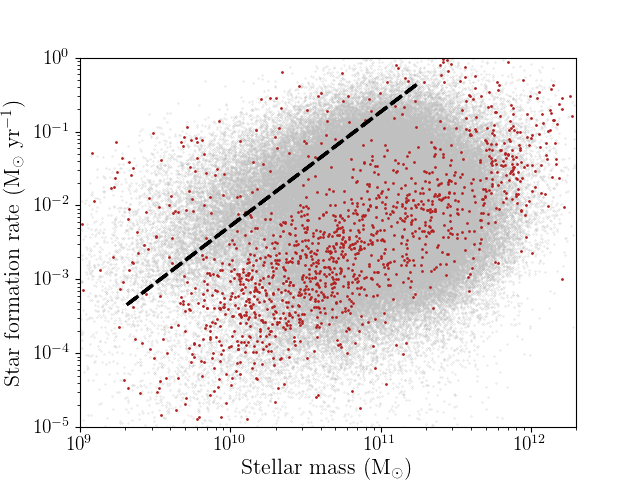}
    \caption{
    SFR as a function of stellar mass.
    Grey dots: all galaxies; red dots: high-$\Sigma_{*}$ galaxies.
    Dashed line: the line fit to the local minimum of the distribution, separating star-forming and passive galaxy sequences.
    }
    \label{fig6}
\end{figure}

We investigate the fundamental properties, colour and SFR, of these high-$\Sigma_*$ galaxies. Figure \ref{fig5} shows the $g-r$ colour as a function of stellar mass.
We find that majority of the high-$\Sigma_{*}$ galaxies show red colours.
The high-$\Sigma_*$ galaxies have the $g-r$ colour on average 0.81 with $\sigma=0.12$.
We take SFR into account for a detailed analysis.
In Figure \ref{fig6}, we found that most of the high-$\Sigma_{*}$ galaxies are passive and have low SFR. 
To emphasise the passive galaxy sequence in Figure \ref{fig6}, we counted the galaxies and calculated the local minimum for each stellar mass bin in Figure \ref{fig6}. We fit the minimum by a line. The line separates the star-forming and passive galaxy sequences in Figure \ref{fig6}, and about 76\% of high-$\Sigma_{*}$ galaxies are located in passive sequence. 

\section{Discussion}
\label{discussion}
\subsection{Maximum density depending on stellar mass}
\cite{Grudic2019} proposed a maximum limit of stellar surface density for any stellar system, spanning $\sim$ 8 order of magnitude in stellar mass from star clusters and galaxies.
The proposed constant maximum value, $\Sigma^{\max}_{*}$, is $\sim 3\times10^{5}$ M$_{\odot}$ pc$^{-2}$.
Galaxies detected with the Hubble Space Telescope are located below the proposed maximum value \citep{van2014,Grudic2019}.
Almost all of our SDSS sample are also well below this value, supporting the existence of the maximum density.
However, Figure \ref{fig3} demonstrates that the maximum stellar surface density depends on stellar mass.
This dependence is also demonstrated in Figure 1 of \cite{Grudic2019} for the sample by \cite{van2014}.
The slope of maximum density is relatively steep in less massive galaxies from log M$_{*}=9$  to 10 $M_{\odot}$ with a flattening or turnover around log M$_{*}$=10.5 $M_{\odot}$.
Obviously less massive galaxies show smaller maximum stellar surface density than massive ones, suggesting that the maximum value is not exactly constant.

\cite{Grudic2019} introduced a constant characteristic surface density, $\Sigma_{\rm crit}$, which parameterises the relative strength of stellar feedback to the self-gravity force in a collapsing cloud.
Here the feedback suppresses star-forming efficiency and self-gravity force enhances it. 
According to their model, star-forming efficiency is determined by the local physics in the collapsing cloud.
The balance between self-gravity force and feedback in the gas cloud determines the efficiency, i.e., $\Sigma/\Sigma_{\rm crit}$ where $\Sigma$ is a total mass surface density of the gas cloud.
As the gas cloud collapses, $\Sigma$ overcomes $\Sigma_{\rm crit}$.
At this point star-forming efficiency is dramatically enhanced and then gas is rapidly consumed. 
As a consequence, star formation terminates before the system becomes extremely dense due to the lack of molecular gas, which places a maximum surface density.
Here a \lq \lq quenching time scale\rq \rq, that is defined by time between reaching $\Sigma_{\rm crit}$ and gas depletion, is a key factor to determine the maximum surface density, because the longer quenching time means that the system has enough time to obtain the higher surface density before the star formation terminates, and vice versa.

Our finding, the stellar-mass dependency of maximum stellar surface density, might suggest a stellar-mass dependency of the quenching time scale. 
One possible physical explanation is an efficiency of stellar feedback.
It is likely that the relative strength of feedback is determined by not only the local physics but also global galaxy potential that is not taken into account in the model by \cite{Grudic2019}. 
The stellar feedback should be more effective in less massive galaxies in which gas cloud is easily swept away by feedback due to the shallow gravitational well \citep[e.g.,][]{Tremonti2004,Chisholm2018}.
In this sense less massive galaxies could have an effectively shorter quenching time scale, during which the galaxies do not have enough time to gain higher stellar surface densities.

Another scenario is a shorter depletion time scale of molecular gas in less massive galaxies.
Actually a positive correlation between the depletion time scale and stellar mass was reported for local galaxies \citep[e.g.,][]{Saintonge2011}.
It is, therefore, possible that less massive galaxies quickly consume remaining molecular gas and then the stellar surface density can not be as large as massive galaxies.
In other words, maximum star-forming efficiency, $\rm \epsilon_{ff}^{max}$ in Figure 5 in \cite{Grudic2019}, could be stellar-mass dependent, although they introduced $\rm \epsilon_{ff}^{max}$ as a free parameter.

In any case, the model by \cite{Grudic2019} is successful in explaining the existence of maximum stellar surface density but it still needs some modifications to understand the stellar-mass dependency of the maximum density.

\subsection{Stellar population of highly dense galaxies}
In the previous section, we demonstrated that the maximum stellar surface density is not constant but dependent on stellar mass, which is probably related to the quenching time scale of star formation.
If the quenching is important, such high surface density galaxies should consist of a red stellar population. 
In Figure \ref{fig5}, we confirmed that high density galaxies occupy the reddest colour distribution in any stellar mass as we expect, except for very massive galaxies from log M$_{*}=$ 11 to 12.
At the massive end it is possible that dusty star-forming galaxies contaminate the red colour distribution. 
Therefore, in Figure \ref{fig6}, we use SFR as a more physical parameter than the apparent colour.
Fig \ref{fig6} clearly shows that a majority of high surface density galaxies are located at a branch of passive galaxies that is below the \lq \lq main sequence\rq \rq\ star-forming galaxies.
This indicates that high density galaxies are already quenched. 

Our results support the idea that galaxy obtains highest surface density as a consequence of galaxy evolution that includes stellar-mass dependent feedback or star-forming efficiency.

\subsection{Surface density of stellar systems less massive than galaxy}
Compared with the other less massive systems shown in Figure 1 of \cite{Grudic2019}, galaxies in our sample have much lower surface density, i.e. 1 dex in $\Sigma_*$.
This is likely due to averaging the global surface density of galaxies over the entire galaxy structures. 
Therefore, less massive galaxies in our sample could contain very dense inner structures (e.g. $\Sigma_{*}\sim 3\times10^{5}$M$_{\odot}$pc$^{-2}$) which do not explicitly appear in the sample.

The global density is likely affected by a global gravitational potential rather than a local gravity.
In contrast, inner structures of a galaxy, such as nuclear star cluster (NSC), 
are likely predominantly controlled by the local gravitational potential rather than the global galaxy potential.
Due to the difference in physical scales, the surface density of the inner structures could be larger the density averaged over the entire galaxy. Therefore, two sequences in Figure 1 of \cite{Grudic2019} of galaxy and inner structures could reflect physics working at different physical scales. Thus, the global surface density of the galaxy, which is of interest in this study, would not be significantly affected by the density of the inner structures.

Another type of stellar system of our concern is ultra-compact dwarfs (UCDs). 
Even though UCDs are galaxies and not inner structures, some previous studies \citep[e.g., ][]{2003MNRAS.344..399B} show that they originate from the stripped cores of once larger galaxies.
Therefore, UCDs can be considered as the former inner structures of galaxies, as in the case of NSCs and others. 
In addition, compact ellipticals (cEs) with M$_{*}$ $\sim10^{8}$M$_{\odot}-10^{9.5}$M$_{\odot}$ show $\Sigma_{*}\sim 10^{3.5}$M$_{\odot}$pc$^{-2}$ \citep{Grudic2019}. 
These galaxies actually follow the dependency of $\Sigma_{*}^{\rm max}$ on stellar mass as shown
in Figure \ref{fig3}. 

\subsection{Density evolution of massive passive galaxies}
In addition to the quenching time, a galaxy density evolution could be another key factor to determine the apparent maximum surface density of our sample.
According to the downsizing galaxy formation scenario, massive galaxies have formed earlier than less massive galaxies.
Massive passive galaxies were found at high redshift Universe up to $z \sim 4$ \citep[e.g.,][]{vanDokkum2008,Kubo2018}.
Such galaxies are dominated by already evolved stellar populations and likely evolve into most massive passive galaxies at $z=0$ without significant star formation, because their stellar mass is already comparable to the massive end of the local galaxies.
Dry minor/major mergers are possible evolution scenarios of massive passive galaxies \citep[e.g.,][]{Bezanson2009,Naab2009}.
The minor merger scenario predicts $\sim$ 2.5 dex evolution in log $\Sigma_{*}$ from $z \sim 4$ to the present for a galaxy with log M$_{*} \sim 11 M_{\odot}$ \citep{Kubo2018}. 
Here surface density monotonically decreases with decreasing redshift.
If this evolution is retroactively applied to the high surface density galaxies with log M$_{*} \sim 11 M_{\odot}$ in our sample, the expected surface density at $z \sim 4$ is $\sim$ $10^{6.5}$ M$_{\odot}$ pc$^{-2}$.
This value is well beyond $\Sigma_{*}^{\rm max}$ proposed by \cite{Hopkins2010,Grudic2019}.
It is suggested that either (i) the proposed value of the maximum density or (ii) minor merger scenario has a difficulty to explain the density evolution of the massive high surface density galaxies in the local Universe.

In the case (i), $\Sigma_{*}^{\rm max}$ could be much higher than the proposed value of 3$\times 10^{5}$ M$_{\odot}$ pc$^{-2}$, because some nuclear star clusters are actually beyond this value \citep{Grudic2019} and the model predicts a higher maximum value ($\sim$ $10^{6.5}$M$_{\odot}$ pc$^{-2}$) when $\epsilon_{ff}^{\rm max}$ is low (e.g.,$\sim$ 0.3).
Some massive galaxies might have lower $\epsilon_{ff}^{\rm max}$ at even high-$z$ Universe as suggested for the local massive galaxies \citep[e.g.,][]{Saintonge2011}.
Therefore some massive passive galaxies could have formed with an extremely high surface density beyond 3$\times 10^{5}$ M$_{\odot}$ pc$^{-2}$ at high-$z$.
Due to the downsizing galaxy formation, such massive galaxies experienced with a longer-term density evolution via merger than less massive ones after the quenching of star formation.
The different formation epoch results in the different degree of density evolution, which might create the flattening or turn over of surface density around log M$_{*}=10.5$ at the local Universe as seen in Figure \ref{fig3}.

The case (ii) suggests that the rapid density evolution by minor merger is not the case of local massive high surface density galaxies.
The major merger changes the surface density mildly because the size increases in proportion to stellar mass \citep[$r \propto $M;][]{Naab2009} in contrast to the rapid size evolution of minor merger \citep[$r \propto$ M$^{2}$;][]{Bezanson2009}.
The 2.5 dex density evolution of the minor merger from $z=4$ to 0 corresponds to only $\sim$ 0.8 dex in the major merger scenario.
The $\sim$ 0.8 dex density evolution is small enough for the local high surface density galaxies to be consistent with the upper limit of 3 $\times 10^{5}$ M$_{\odot}$ pc$^{-2}$.

In either case, investigating the extremely dense galaxies at high-z Universe is a key to address the density evolution of high surface density galaxies and the existence of the universal maximum limit on the surface density.

\subsection{PSF effect on surface density}
Here we discuss a possible impact of point spread function (PSF) on the surface stellar density estimate. 
We used the size measurement deconvoluted with PSF function to calculate the stellar surface density.
This is ideally free from PSF broadening.
However, in practice, the deconvolution is unlikely to be accurate when the apparent size of galaxy is comparable to the PSF size.

\begin{figure}
	\includegraphics[width=.52\textwidth]{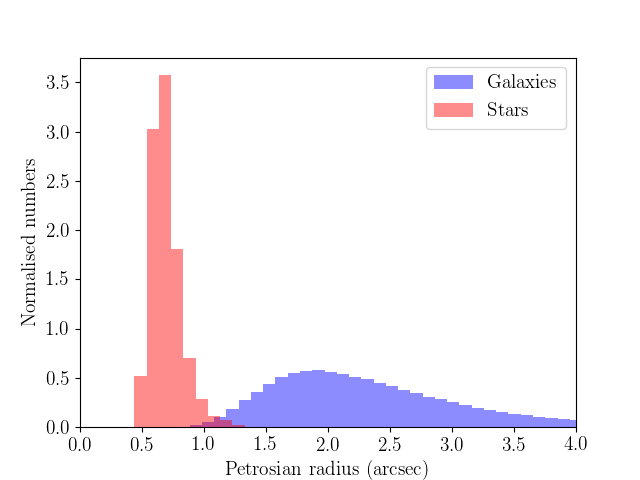}
    \caption{
    Histograms of apparent sizes of stars and the galaxy sample from the whole SDSS catalogue.
    }
    \label{fig7}
\end{figure}

Figure \ref{fig7} shows the histograms of apparent sizes of stars and all galaxies in the whole SDSS catalogue.
Stars are selected by specifying the SDSS object type, i.e., type=Star, with a magnitude cut of m$_{r}$ $<$ 17.77 mag to exclude too faint stars.
We used CLEAN flag to exclude saturated stars. 
In Figure \ref{fig7}, the apparent size of stars peaks at $\sim$ 0$\arcsec$.65 that is well below the majority of galaxy sample.
Although two histograms are slightly overlapped at the edges of their distributions, the number of overlapped galaxies is much smaller (only $\sim4\%$) than the whole galaxy sample.
Therefore we suggest that there is no significant impact of PSF on estimating the stellar surface density in our sample.

\section{Conclusion}
\label{conclusion}
In this study, we compile a large sample of local galaxies ($0.05<z<0.5$) from the SDSS DR12 to investigate what physics determines the maximum stellar-surface density,  $\Sigma^{\max}_{*}$.
Our findings are summarised as follows.

\begin{enumerate}
\item We find that maximum stellar surface density, $\Sigma^{\max}_{*}$, of galaxies depends on stellar mass, and is not a constant as predicted by previous theoretical models.

\item We find the stellar mass dependence of $\Sigma^{\max}_{*}$ is stronger for less massive galaxies with $\sim10^{10}$M\textsubscript{\(\odot\)} or less (Figure \ref{fig3}). One possible interpretation of the stellar-mass dependency of $\Sigma^{\max}_{*}$ is mass-dependent stellar feedback.

\item We find that high-$\Sigma_{*}$ galaxies have already quenched star formation. 
In Figure \ref{fig5}, the high-$\Sigma_{*}$ galaxies are red, i.e., $g-r\sim$0.8 on average.
In Figure \ref{fig6}, a majority of the high-$\Sigma_{*}$ galaxies are located at the passive branch of galaxies, where their SFRs are lower than those of the star-forming galaxies in the \lq \lq main sequence\rq \rq. 
Our results support the idea that galaxy obtains highest surface
density as a consequence of galaxy evolution that includes stellar-mass dependent feedback.

While we focus on presenting the observational results in this study, we urge theorists to reproduce our results with simulations and test our qualitative interpretations quantitatively in future works.
\end{enumerate}

\section*{Acknowledgements}
We are very grateful to the anonymous referee for many insightful comments.
TG acknowledges the supports by the Ministry of Science and Technology of Taiwan through grants 105-2112-M-007-003-MY3 and 108-2628-M-007-004-MY3.
TH and AYLO are supported by the Centre for Informatics and Computation in Astronomy (CICA) at National Tsing Hua University (NTHU) through a grant from the Ministry of Education of the Republic of China (Taiwan).
AYLO's visit to NTHU is also supported by the Ministry of Science and Technology of the ROC (Taiwan) grant 105-2119-M-007-028-MY3, kindly hosted by Prof. Albert Kong.

\bibliographystyle{mnras}
\bibliography{REF} 

\bsp	
\label{lastpage}
\end{document}